\documentclass[11pt]{article}

\usepackage[english]{babel}
\usepackage[tbtags]{amsmath}
\usepackage{amssymb}
\usepackage{mathtools}
\usepackage[affil-it,auth-sc]{authblk}
\usepackage{geometry}
\usepackage{graphicx}
\usepackage{hyperref}
\usepackage{cleveref}
\usepackage{xcolor}
\usepackage{tocloft}
\usepackage{enumitem}

\usepackage[style=numeric-comp,sorting=none,firstinits=true,doi=false]{biblatex}
\DeclareFieldFormat[article]{title}{\textit{#1}}
\DeclareFieldFormat[article]{volume}{\textbf{#1}}

\newcommand{\mleft}{\mathopen{}\mathclose\bgroup\left}
\newcommand{\mright}{\aftergroup\egroup\right}
\renewcommand{\Re}[2][]{\operatorname{Re}\mleft(#2\mright)^{#1}}
\renewcommand{\Im}[2][]{\operatorname{Im}\mleft(#2\mright)^{#1}}

\newcommand{\QB}{Q_\mathrm{B}}
\newcommand{\imaginaryi}{\mathrm{i}}
\newcommand{\Psidag}{{\Psi^\ddagger}}
\newcommand{\eulere}[1]{\mathrm{e}^{#1}}
\renewcommand{\d}{\mathrm{d}}
\newcommand{\integlbndry}[3][]{\int\limits_{#2}^{#1}{\!\d#3}\;}	
\newcommand*{\defeq}{\mathrel{\vcenter{\baselineskip0.5ex \lineskiplimit0pt\hbox{\scriptsize.}\hbox{\scriptsize.}}}=}

\newcommand{\MoveEqLeftFactor}{1.5}
\newcommand{\MultlinedFactor}{0.80}

\newcommand{\tr}{\mathrm{tr}}

\DeclareCiteCommand{\supercite}[\mkbibsuperscript]
{\usebibmacro{cite:init}%
	
	\iffieldundef{prenote}
	{}
	{\BibliographyWarning{Ignoring prenote argument}}%
	\iffieldundef{postnote}
	{}
	{\BibliographyWarning{Ignoring postnote argument}}%
	\bibopenbracket}%
{\usebibmacro{citeindex}%
	\usebibmacro{cite:comp}}
{}
{\usebibmacro{cite:dump}\bibclosebracket}

\title{\textbf{Real Tachyon Vacuum Solution\\ without Square Roots}}
\author[1]{Maximilian Jokel\thanks{\url{maximilian.jokel@tum.de}}}
\affil[1]{\small{Arnold Sommerfeld Center for Theoretical Physics\protect\\ Ludwig-Maximilians-Universit\"at\protect\\ Theresienstra\ss e 37\protect\\ 80333 M\"unchen, Germany}}

\begin{document}

\allowdisplaybreaks
\setlength{\affilsep}{1cm}		
\numberwithin{equation}{section}



\setlength{\abovedisplayskip}{14pt plus 3pt minus 9pt}
\setlength{\belowdisplayskip}{14pt plus 3pt minus 9pt}
\setlength{\jot}{4pt}										


\maketitle

\begin{abstract}
\noindent
In this note we present a real tachyon vacuum solution for open bosonic string field theory which does not require computing square roots of states in the wedge algebra. We study its algebraic structure, prove the equation of motion and show that the solution supports no cohomology. Finally we show that its energy is in agreement with the value predicted by Sen's conjecture.
\end{abstract}


\newpage
\tableofcontents
\vspace*{15pt}
\section{Introduction}\label{sec:Introduction}
One of the main ingredients of string field theory is the requirement that the action should be real. Customarily and most effectively, this requirement is implemented by demanding that string fields obey the reality condition
	\begin{equation}\label{eq:string_field_reality_condition}
		\Phi^\ddagger=\Phi
	\end{equation}
where the double dagger operation denotes a composition of Hermitian and BPZ conjugation as described by Gaberdiel and Zwiebach~\supercite{Gaberdiel1997}. Although the reality condition is sufficient to guarantee that the action is real, it is not necessary. Oftentimes the simplest analytic solutions are not real. Examples include the Erler-Schnabl tachyon vacuum solution~\supercite{Erler2009Simple}, the Kiermaier-Okawa solution for marginal deformations~\supercite{Kiermaier2007jul,Fuchs2007marg}, and many solutions in the Berkovits open superstring field theory~\supercite{Erler2013,Erler2007,Fuchs2007supermarg,Okawa2007,Kiermaier2007,Noumi2011}. Although these solutions fail to satisfy the reality condition, they are acceptable since they can be shown to be gauge equivalent to real solutions. Nevertheless, it is desirable to have analytic solutions whose reality is manifest. Consider for example the Erler-Schnabl solution which in its simplest formulation takes the non-real form
	\begin{equation}
		\Psi=c(1+K)Bc\frac{1}{1+K}\label{eq:Erler_Schnabl_solution_non_real_form}
	\end{equation}
By similarity transformation one may find an equivalent real solution
	\begin{equation}
		\Psi=\frac{1}{\sqrt{1+K}}c(1+K)Bc\frac{1}{\sqrt{1+K}}\label{eq:Erler_Schnabl_solution_real_form}
	\end{equation}
at the cost of introducing somewhat awkward square roots. Similar and oftentimes more complicated square roots appear in real solutions of various kinds throughout the literature. It is interesting to look for alternatives with a different kind of structure.\newline\newline\newline
In this note we show that given a tachyon vacuum solution $\Psi$ along with a homotopy operator $A$~\supercite{Ellwood2001,Ellwood2006} satisfying
	\begin{equation}
		A^2=0\qquad\text{and}\qquad A^\ddagger=A\label{eq:assumptions_homotopy_operator}
	\end{equation}
the following expression
	\begin{equation}
		\Phi=\Re{\Psi}+\Im{\Psi}A\Im{\Psi}\label{eq:new_real_solution}
	\end{equation}
is a real solution for the tachyon vacuum. Applied to the Erler-Schnabl solution \eqref{eq:Erler_Schnabl_solution_non_real_form}, the real tachyon vacuum takes the form
	\begin{equation}
		\Phi=c-\frac{1}{2}\bigg(c\partial c\frac{B}{1+K}+\frac{B}{1+K}c\partial c\bigg)-\frac{1}{4}\frac{B}{1+K}c\partial c\frac{B}{1+K}c\partial c\frac{B}{1+K}\label{eq:new_real_solution_Erler_Schnabl}
	\end{equation}
which does no longer involve square roots. Interestingly, the solution does not take the factorized form
	\begin{equation}
		\Psi=Fc\frac{KB}{1-FG}cG\label{eq:factorized_form}
	\end{equation}
widely studied throughout the literature. It would be worth knowing to see if there is a generalization which gives real solutions for other backgrounds, for example along the lines of the work on background-independent solutions by Erler and Maccaferri.\newline\newline
This article is organized as follows: To begin with, in \autoref{sec:Algebraic_Structure_of_the_Solution} we study the algebraic structure of the solution. We prove that the solution satisfies the equation of motion and show that it represents the tachyon vacuum by demonstrating the absence of cohomology. We continue by deriving the finite gauge transformation relating $\Psi$ to $\Phi$ from an exact left gauge transformation~\supercite{Erler2012}. Further on, in \autoref{sec:action} we show that the action of $\Phi$ is the same as the action of $\Psi$. In \autoref{sec:Solution_in_the_KBc_subalgebra} we restrict our attention to the $K\!Bc$ subalgebra and derive the $K\!Bc$ form of the new solution. We apply the result to the Erler-Schnabl solution, compute the energy of the solution in the $K\!Bc$ framework and show that it is in agreement with the value predicted by Sen's conjectures. Finally, we confirm that the solution does not disturb the closed string background by demonstrating that the closed string tadpole vanishes.

\newpage
\section{Algebraic Structure of the Solution}\label{sec:Algebraic_Structure_of_the_Solution}
To begin with, we shall recall some basic properties of reality conjugation introduced in \eqref{eq:string_field_reality_condition}. The double dagger is an involution of the star algebra and satisfies the following basic properties
	\begin{subequations}\label{eq:properties_reality_conjugation}
	\begin{align}
		(\Psi_1\Psi_2)^\ddagger&=\Psi_2^\ddagger\Psi_1^\ddagger\\
		(\QB\Psi)^\ddagger&=(-1)^{\varepsilon(A)+1}\QB(\Psi^\ddagger)\\
		\tr[\Psi]^\ast&=\tr[\Psi^\ddagger]
	\end{align}
	\end{subequations}
Here and in what follows, $(\;\cdot\;)^\ast$ and $\varepsilon(\;\cdot\;)$ denote complex conjugation and Grassmannality, respectively.\newline\newline\noindent
We start by proving the equation of motion for the proposed solution. The proof is done efficiently by using the following relations\footnote{Here and in the following, $\big[\;\cdot\;,\;\cdot\;\big]$ denotes the graded commutator, defined as
	\begin{align*}
		\big[A,B\big]\defeq AB-(-1)^{\varepsilon(A)\varepsilon(B)}BA
	\end{align*}
with $\varepsilon(A)$ encoding the Grassmannality of the string field $A$.}
	\begin{subequations}
		\begin{align}
			\big[\!\!\:\Re{\Psi},\Im{\Psi}\!\:\big]&=\Im{\Psi^2}\label{eq:relation_1}\\
			\Re[2]{\Psi}-\Im[2]{\Psi}\vphantom{\big]}&=\Re{\Psi^2}\label{eq:relation_2}\\
			Q_{\Re{\Psi}}A\vphantom{\big]}&=1\phantom{\Im{\Psi^2}}\label{eq:relation_3}\\
			\big[\!\!\:\Im{\Psi},A\!\:\big]&=0\label{eq:relation_4}
		\end{align}
	\end{subequations}
The first two relations can be immediately verified by inserting the definitions of $\Re{\Psi}$ and $\Im{\Psi}$. The last two relations are symmetric and antisymmetric linear combinations of the relations $Q_\Psi A=1$ and $Q_\Psidag A=1$, respectively. The latter hold since both $\Psi$ and $\Psidag$ are tachyon vacuum solutions by assumption.\newline\newline
With these relations in mind, we consider the kinetic and potential terms separately one after the other. Using the derivation properties of $\QB$ along with \eqref{eq:relation_3}, the kinetic term can be rewritten as
	\begin{align}
		\QB\Phi&=\QB\Re{\Psi}+\QB\Big(\Im{\Psi}A\Im{\Psi}\Big)\\
		&\begin{multlined}[b][0.75\textwidth]
			\!=\QB\Re{\Psi}-\Im{\Psi}\Big(\QB A\Big)\Im{\Psi}+\\[4pt]
			+\Big(\QB\Im{\Psi}\Big)A\Im{\Psi}+\Im{\Psi}A\Big(\QB\Im{\Psi}\Big)
		\end{multlined}\\
		&\begin{multlined}[b][0.75\textwidth]
			\!\overset{\mathclap{\scalebox{0.5}{\eqref{eq:relation_3}}}}{=}-\Re{\Psi^2}-\Im{\Psi}\Big(1-\big[\!\!\:\Re{\Psi},A\!\:\big]\Big)\Im{\Psi}+\\[4pt]
			-\Im{\Psi^2}A\Im{\Psi}-\Im{\Psi}A\Im{\Psi^2}
		\end{multlined}\\
		&\begin{multlined}[b][0.75\textwidth]
			\!=-\Big(\Re{\Psi^2}+\Im[2]{\Psi}\Big)+\Big(\Im{\Psi}\Re{\Psi}-\Im{\Psi^2}\Big)A\Im{\Psi}+\\[4pt]
			+\Im{\Psi}A\Big(\Re{\Psi}\Im{\Psi}-\Im{\Psi^2}\Big)
		\end{multlined}\\
		&\overset{\mathclap{\scalebox{0.5}{\eqref{eq:relation_1}}}}{=}-\Big(\Re{\Psi^2}+\Im[2]{\Psi}\Big)-\big[\!\!\:\Re{\Psi},\Im{\Psi}A\Im{\Psi}\!\:\big]
	\end{align}
In the next-to-last step we just grouped terms suitably to see that \eqref{eq:relation_1} helps to simplify the expression further. Finally, for the last equality sign we exploited \eqref{eq:relation_4} to combine\nopagebreak[4] the latter two parentheses into one graded commutator.\newline\newline
For the potential term, things are even simpler than for the kinetic term. Taking the square of $\Phi$ and grouping terms appropriately, we find
	\begin{align}
		\Phi^2&=\big(\Re{\Psi}+\Im{\Psi}A\Im{\Psi}\big)^2\\[2.0mm]		
		&=\Re[2]{\Psi}+\big[\!\!\:\Re{\Psi},\Im{\Psi}A\Im{\Psi}\!\:\big]+\Im{\Psi}A\Im[2]{\Psi}A\Im{\Psi}\\[2.0mm]
		&\overset{\mathclap{\scalebox{0.5}{\eqref{eq:relation_4}}}}{=}\Re[2]{\Psi}+\big[\!\!\:\Re{\Psi},\Im{\Psi}A\Im{\Psi}\!\:\big]
	\end{align}
In the last step we exploited \eqref{eq:relation_4} to commute the homotopy operators $A$ adjacent to each other and therefore make the last term vanish due to the assumed nilpotency of the homotopy operator $A$. Adding up the kinetic and potential term, the graded commutators cancel and we are left with
	\begin{equation}
		\QB\Phi+\Phi^2=-\Re{\Psi^2}+\Re[2]{\Psi}-\Im[2]{\Psi}\overset{\scalebox{0.5}{\eqref{eq:relation_2}}}{=}0\label{eq:equation_of_motion_Phi}
	\end{equation}
which completes the proof of the equation of motion.\newline\newline\noindent
Having established the equation of motion for the new solution, we next have to prove that $\Phi$ is a tachyon vacuum solution. For this purpose, it is sufficient to show that $\Phi$ supports no open string states. This can be seen as follows:
	\begin{align}
		Q_\Phi A&=\QB A+\big[\Phi,A\!\:\big]\\
		&\overset{\mathclap{\scalebox{0.5}{\eqref{eq:new_real_solution}}}}{=}\QB A+\Re{\Psi}A+A\Re{\Psi}+A\Im{\Psi}A\Im{\Psi}+\Im{\Psi}A\Im{\Psi}A\\
		&\overset{\mathclap{\scalebox{0.5}{\eqref{eq:relation_4}}}}{=}\frac{1}{2}\big(\QB A+A\Psi+\Psi A\big)+\frac{1}{2}\big(\QB A+A\Psidag+\Psidag A\big)\\
		&=1
	\end{align}
Since $A$ is a homotopy operator for $Q_\Phi$, the cohomology is empty and the solution supports no open string states.\newline\newline\noindent
Next we want to study the connection between the real solution $\Phi$ and the non-real solution $\Psi$ we started with. To find a gauge transformation relating them, it is helpful to recall the work done by Erler and Maccaferri~\supercite{Erler2012}. They show that any pair of solutions $\Phi_1,\Phi_2$ can be related by a so-called \emph{left gauge transformation} which is a ghost number zero string field $U$ satisfying
	\begin{equation}
		Q_{\Phi_1\!\Phi_2}U=0\label{eq:definition_left_gauge_transformation}
	\end{equation}
where $Q_{\Phi_1\!\Phi_2}$ defined as
	\begin{equation}
		Q_{\Phi_1\!\Phi_2}(\;\cdot\;)\defeq\QB(\;\cdot\;)+\Phi_1(\;\cdot\;)-(-1)^{\varepsilon(\;\cdot\;)}(\;\cdot\;)\Phi_2\label{eq:definition_nilpotent_operator}
	\end{equation}
is the kinetic operator for a stretched string connecting the classical solutions $\Phi_1$ and $\Phi_2$. If $U$ happens to have a star algebra inverse, then \eqref{eq:definition_left_gauge_transformation} implies that the solutions $\Phi_1$ and $\Phi_2$ are gauge equivalent
	\begin{equation}
		U^{-1}\big(\QB+\Phi_1\big)U=\Phi_2
	\end{equation}
An important class of left gauge transformations are \emph{exact left gauge transformations} which take the form
	\begin{equation}
		U=Q_{\Phi_1\!\Phi_2}b
	\end{equation}
where $b$ is a string field at ghost number $\#_\text{gh}(b)=-1$. It is an elementary result that an exact left gauge transformation can only have a star algebra inverse if $\Phi_1$ and $\Phi_2$ are tachyon vacuum solutions.\newline\newline\noindent
Applying this result to our case, we can postulate that $\Psi$ and $\Phi$ as tachyon vacuum solutions are gauge equivalent, and the explicit form of the finite gauge parameter can be found in the form of an exact left gauge transformation. Since the only natural string field at ghost number $-1$ in our setup is the homotopy operator $A$, it is natural to consider an exact left gauge transformation of the form
	\begin{subequations}\label{eq:complete_gauge_transformation}
	\begin{align}
		U&=Q_{\Phi\!\!\;\Psi}A=1-\imaginaryi A\Im{\Psi}\label{eq:gauge_transformation}
\intertext{It is easy to check that $U$ has an inverse}
		U^{-1}&=1+\imaginaryi A\Im{\Psi}\label{eq:inverse_gauge_transformation}
	\end{align}
	\end{subequations}
and therefore defines a finite gauge transformation relating $\Psi$ to $\Phi$.

\section{Action}\label{sec:action}
We now show that action evaluated on the new real solution $\Phi$ is the same as the action evaluated on $\Psi$. We accomplish this by showing that the action is preserved by the finite gauge transformation \eqref{eq:complete_gauge_transformation}.\footnote{The Chern-Simons action is not identically invariant under finite gauge transformations, due to the possibility of large gauge transformations. A proposed example in string field theory is the work by Hata and Kojita~\supercite{Hata2011} on multibrane solutions.}\newline\newline\noindent
Using the equation of motion for $\Psi$, the action be rewritten as
	\begin{equation}
		\mathcal{S}[\Psi]=\frac{1}{2}\tr\big[\Psi\QB\Psi\big]+\frac{1}{3}\tr\big[\Psi^3\big]=\frac{1}{6}\tr\big[\Psi^3\big]
	\end{equation}
We want to confirm that the action $\mathcal{S}[\Phi]$ for the new solution $\Phi$ reduces to $\mathcal{S}[\Psi]$. Inserting
	\begin{equation}
		\Phi=U^{-1}\big(\QB+\Psi\big)U
	\end{equation}
into the above expression, multiplying out all the parentheses and regrouping the resulting terms suitably, we find after some straightforward manipulations
	\begin{equation}
		\mathcal{S}[\Phi]=\frac{1}{6}\tr\bigg[\Big((\QB U)U^{-1}\Big)^3+3\Big((\QB U)U^{-1}\Big)^2\Psi+3\Big((\QB U)U^{-1}\Big)\Psi^2+\Psi^3\bigg]\label{eq:EPhi}
	\end{equation}
In order to show that $\mathcal{S}[\Phi]=\mathcal{S}[\Psi]$, all the terms involving $U$ and $U^{-1}$ have to vanish identically. Using the equations of motion for the pure-gauge solution $-(\QB U)U^{-1}$ and the solution $\Psi$, we arrive at
	\begin{align}
		\mathcal{S}[\Phi]&=\frac{1}{6}\tr\bigg[\Big((\QB U)U^{-1}\Big)^3+3\QB\Big((\QB U)U^{-1}\Big)\Psi-3\Big((\QB U)U^{-1}\Big)\QB\Psi+\Psi^3\bigg]\\[4pt]
		&=\frac{1}{6}\tr\bigg[\Big((\QB U)U^{-1}\Big)^3+3\QB\Big((\QB U)U^{-1}\Psi\Big)+\Psi^3\bigg]\label{eq:energy_calculation_intermediate_result_1}
	\end{align}
where the resulting $\QB$-exact term does not contribute to the action. To show that the first term also vanishes, we compute $(\QB U)U^{-1}$ using \eqref{eq:gauge_transformation} and \eqref{eq:inverse_gauge_transformation} and find
	\begin{align}
		(\QB U)U^{-1}&=-\QB\big(\imaginaryi A\Im{\Psi}\big)\Big(1+\imaginaryi A\Im{\Psi}\big)\label{eq:QBU_inverseU}
	\end{align}
According to \eqref{eq:relation_4} the string field $\Im{\Psi}$ satisfies $[\!\:\Im{\Psi},A\!\:]=0$ implying 
	\begin{equation}
		\QB(A\Im{\Psi})A\Im{\Psi}=-A\Im{\Psi}\QB(A\Im{\Psi})\label{eq:QB_ImPsi_A}
	\end{equation}
via $\QB(A\Im{\Psi}A\Im{\Psi})=0$. Using these relations along with the equation of motion for the pure-gauge solution $-(\QB U)U^{-1}$, the first term in \eqref{eq:energy_calculation_intermediate_result_1} also reduces to a $\QB$-exact quantity
	\begin{align}
		\MoveEqLeft[\MoveEqLeftFactor]\tr\bigg[\Big((\QB U)U^{-1}\Big)^3\bigg]\nonumber\\[2pt]
		&=\tr\bigg[\QB\Big((\QB U)U^{-1}\Big)\Big((\QB U)U^{-1}\Big)\bigg]\\[2pt]
		&\overset{\mathclap{\scalebox{0.5}{\eqref{eq:QBU_inverseU}}}}{=}\tr\bigg[\Big(\QB(\imaginaryi A\Im{\Psi})\Big)^2\Big(\QB\big(\imaginaryi A\Im{\Psi})(1+\imaginaryi A\Im{\Psi})\Big)\bigg]\\[2pt]
		&\begin{multlined}[b][\MultlinedFactor\textwidth]
			\!=\tr\bigg[\Big(\QB\big(\imaginaryi A\Im{\Psi}\big)\Big)^3\bigg]+\frac{1}{2}\tr\bigg[\Big(\QB\big(\imaginaryi A\Im{\Psi}\big)\Big)^3\Big(\imaginaryi A\Im{\Psi}\Big)\bigg]+\\[6pt]
			+\frac{1}{2}\tr\bigg[\Big(\QB\big(\imaginaryi A\Im{\Psi}\big)\Big)^3\Big(\imaginaryi A\Im{\Psi}\Big)\bigg]
		\end{multlined}\\[2pt]
		&\begin{multlined}[b][\MultlinedFactor\textwidth]
			\!=\tr\bigg[\Big(\QB\big(\imaginaryi A\Im{\Psi}\big)\Big)^3\bigg]+\frac{1}{2}\tr\bigg[\Big(\QB\big(\imaginaryi A\Im{\Psi}\big)\Big)^3\Big(\imaginaryi A\Im{\Psi}\Big)\bigg]-\\[6pt]
			-\frac{1}{2}\tr\bigg[\Big(\QB\big(\imaginaryi A\Im{\Psi}\big)\Big)^2\Big(\imaginaryi A\Im{\Psi}\Big)\Big(\QB\big(\imaginaryi A\Im{\Psi}\big)\Big)\bigg]
		\end{multlined}\\[2pt]
		&\begin{multlined}[b][\MultlinedFactor\textwidth]
			\!=\tr\bigg[\Big(\QB\big(\imaginaryi A\Im{\Psi}\big)\Big)^3\bigg]+\frac{1}{2}\tr\bigg[\Big(\QB\big(\imaginaryi A\Im{\Psi}\big)\Big)^3\Big(\imaginaryi A\Im{\Psi}\Big)\bigg]-\\[6pt]
			-\frac{1}{2}\tr\bigg[\Big(\QB\big(\imaginaryi A\Im{\Psi}\big)\Big)^3\Big(\imaginaryi A\Im{\Psi}\Big)\bigg]
		\end{multlined}\\
		&=\tr\big[\QB(\dots)\big]=0
	\end{align}
For the third-last equality we used \eqref{eq:QB_ImPsi_A} while for the next-to-last equality we exploited the cyclicity of the trace. This completes the proof that all terms in \eqref{eq:energy_calculation_intermediate_result_1} except for the last one vanish and therefore implies that the action $\mathcal{S}[\Phi]$ correctly reduces to $\mathcal{S}[\Psi]$. Similarly it is possible to show that the closed string overlap~\supercite{Ellwood2008} evaluated for $\Psi$ is the same for $\Phi$.

\section{Realization in the $K\!Bc$ Subalgebra}\label{sec:Solution_in_the_KBc_subalgebra}
Finally, we want to study how the proposed real solution is realized in the $K\!Bc$ subalgebra. It can be shown that all tachyon vacuum solutions in the $K\!Bc$ subalgebra take the form~\supercite{Erler2017}
	\begin{equation}
		\Psi=T\frac{KB}{1-F}T+\QB(BT)\label{eq:KBc_solution}
	\end{equation}
where $T$ is the \emph{zero momentum tachyon state} characterizing the solution. It is an arbitrary ghost number one string field satisfying the graded commutator relation
	\begin{equation}
		\big[B,T\!\:\big]=F
	\end{equation}
and $F$ is a function of the string field $K$ satisfying\footnote{In the current understanding, additional conditions on $F$ include:
	\begin{enumerate}[label=(\alph*),leftmargin=15.0mm]
		\item $F(K)<1\text{ for }K>0$
		\item $F(\infty)=0$~\supercite{Erler2012LevelExp}
		\item $F$ must have an inverse Laplace transform~\supercite{Murata2011}
	\end{enumerate}}
	\begin{equation}
		F(0)=1\qquad\text{and}\qquad F'(0)\neq 0
	\end{equation}
Furthermore, the solution possesses a homotopy operator $A$ which is given by
	\begin{equation}
		A=\frac{1-F}{K}B\label{eq:homotopy_operator_KBc}
	\end{equation}
The form \eqref{eq:KBc_solution} can be easily derived from Okawa's pure gauge solution~\supercite{Okawa2006}
	\begin{equation}
		\Psi=V\QB V^{-1}\qquad\text{where}\qquad V=1-BT
	\end{equation}
by expanding $V^{-1}$ as a formal power series, recognizing that $\QB\big((BT)^n\big)=F^{n-1}\QB(BT)$ and exploiting the above graded commutator relation repeatedly. When $T$ is the Fock space zero momentum tachyon state
	\begin{equation}
		T=\sqrt{\Omega}c\sqrt{\Omega}=\frac{2}{\pi}c_1|0\rangle
	\end{equation}
we recover Schnabl's solution~\supercite{Schnabl2005} while choosing
	\begin{align}
		T=c\frac{1}{1+K}
	\end{align}
leads to the Erler-Schnabl solution. For generic ghost number one string fields $T$, the solution \eqref{eq:KBc_solution} does not take a factorized form \eqref{eq:factorized_form} as is usually assumed throughout the literature~\supercite{Schnabl2005,Okawa2006}.\newline\newline\noindent
In order to establish the connection between the above tachyon vacuum solution $\Psi$ and the new real solution $\Phi$, recall that
	\begin{equation}
		K^\ddagger=K\qquad\qquad B^\ddagger=B\qquad\qquad c^\ddagger=c
	\end{equation}
Since we have assumed the homotopy operator to be a real string field in the very beginning, \eqref{eq:homotopy_operator_KBc} implies that $F$ must be a real function of $K$. Inserting \eqref{eq:KBc_solution} into \eqref{eq:new_real_solution} and using the graded-commutator relation $[B,\Re{T}]=F$, we end up with
	\begin{equation}
		\Phi=\Re{T}\frac{KB}{1-F}\Re{T}+\QB\big(B\Re{T}\big)\label{eq:KBc_solution_real}
	\end{equation}
after a lengthy series of basically straightforward manipulations. Therefore, the effect of passing to the real solution $\Phi$ is to replace the zero momentum tachyon state $T$ with its real part.\newline\newline\noindent
As a concrete example, we want to apply the new real solution to the Erler-Schnabl solution
	\begin{equation}
		\Psi=c(1+K)Bc\frac{1}{1+K}\label{eq:Erler_Schnabl_solution}
	\end{equation}
and compute its energy. Exploiting the relations $[B,c\!\:]=1$ and $\QB c=cKc=c\partial c$, the solution takes the form
	\begin{equation}
		\Phi=c-\frac{1}{2}\bigg(c\partial c\frac{B}{1+K}+\frac{B}{1+K}c\partial c\bigg)-\frac{1}{4}\frac{B}{1+K}c\partial c\frac{B}{1+K}c\partial c\frac{B}{1+K}\label{eq:Phi_Erler_Schnabl}
	\end{equation}
A particularly useful form for computing the energy is
	\begin{equation}
		\Phi=\frac{1}{4}\bigg(\frac{1}{1+K}c+c\frac{1}{1+K}+c\frac{B}{1+K}c+\frac{1}{1+K}c\frac{1}{1+K}\bigg)+\text{$\QB$-exact terms}\label{eq:Erler_Schnabl_solution_new_real}
	\end{equation}
since the $\QB$-exact terms will give no contribution to the energy. Inserting this form of the solution into
	\begin{equation}
		E(\Phi)=-\mathcal{S}[\Phi]=+\frac{1}{6}\tr\big[\Phi\QB\Phi\big]
	\end{equation}
and simplifying algebraically as far as possible, we are left with
	\begin{align}
		\begin{multlined}[b][\MultlinedFactor\textwidth]
			E=\frac{1}{96}\tr\bigg[\bigg(\frac{2}{1+K}c\frac{1}{1+K}+\frac{3}{(1+K)^2}c\frac{1}{1+K}+\frac{2}{1+K}c\frac{1}{(1+K)^2}+\\[6pt]
			+\frac{1}{(1+K)^2}c\frac{1}{(1+K)^2}\bigg)cKc-Kc\frac{1}{1+K}c\frac{1}{1+K}c\frac{1}{1+K}cB\bigg]
		\end{multlined}
	\end{align}
Expanding the $K$-dependent factors in terms of wedge states $\Omega=\eulere{-K}$ via the Laplace transform according to
	\begin{align}
		\frac{1}{(1+K)^n}=\integlbndry[\infty]{0}{t}\frac{t^{-1+n}\eulere{-t}}{\Gamma(n)}\eulere{-tK}=\integlbndry[\infty]{0}{t}\frac{t^{-1+n}\eulere{-t}}{\Gamma(n)}\Omega^{t}\label{eq:Laplace_transform}
	\end{align}
the last term involves three integrations while all the other terms only give two integrals. Evaluating the occurring terms separately, we find
	\begin{subequations}
	\begin{align}
		\tr\bigg[\frac{1}{1+K}c\frac{1}{1+K}cKc\bigg]&=-\frac{3}{\pi^2}\\[6pt]
		\tr\bigg[\frac{1}{(1+K)^2}c\frac{1}{1+K}cKc\bigg]&=-\frac{6}{\pi^2}\\[6pt]
		\tr\bigg[\frac{1}{1+K}c\frac{1}{(1+K)^2}cKc\bigg]&=-\frac{6}{\pi^2}\\[6pt]
		\tr\bigg[\frac{1}{(1+K)^2}c\frac{1}{(1+K)^2}cKc\bigg]&=-\frac{10}{\pi^2}-\frac{30}{\pi^4}\\[6pt]
		\tr\bigg[Kc\frac{1}{1+K}c\frac{1}{1+K}c\frac{1}{1+K}cB\bigg]&=\frac{2}{\pi^2}-\frac{30}{\pi^4}
	\end{align}
	\end{subequations}
Adding up all the terms, the energy correctly reduces to the value
	\begin{equation}
		E(\Phi)=-\frac{1}{2\pi^2}
	\end{equation}
in accordance with Sen's conjecture.\newline\newline\noindent

\subsection*{Acknowledgement}
I thank T. Erler for guidance and supervision during the preparation of this brief research note. I also would like to thank Prof. Ivo Sachs for hospitality at Ludwig-Maximilians-Universit\"at M\"unchen where the solution was developed.

\newpage
\printbibliography

\end{document}